\documentclass{aastex62}
\usepackage{mathtools}
\usepackage{amsmath}
\usepackage{listings}

\shorttitle{Pointing Limits of Light Curve Characterization}
\shortauthors{Saunders et al.}

\begin{document}

\title{The Pointing Limits of Transiting Exoplanet Light Curve Characterization with Pixel Level De-correlation}

\correspondingauthor{Nicholas Saunders}
\email{nicholas.k.saunders@nasa.gov}

\author{Nicholas Saunders}
\affil{NASA Ames Research Center, Moffett Field, CA}
\affil{Bay Area Environmental Research Institute, Petaluma, CA}

\author{Rodrigo Luger}
\altaffiliation{Flatiorn Fellow}
\affil{Center for Computational Astrophysics, Flatiron Institute, New York, NY}

\author{Rory Barnes}
\affil{Department of Astronomy, University of Washington, Seattle, WA}
\affil{Virtual Planetary Laboratory, Seattle, WA}

\begin{abstract}

	We present \texttt{scope} (Simulated CCD Observations for Photometric Experimentation), a \texttt{Python} package to create a forward model of telescope detectors and simulate stellar targets with motion relative to the CCD. The primary application of this package is the simulation of the \textit{Kepler} Space Telescope detector to predict and characterize increased instrumental noise in the spacecraft's final campaigns of observation. As the fuel powering the spacecraft's stabilizing thrusters ran out and thruster fires began to sputter and fail, stellar Point Spread Functions (PSFs) experienced more extreme and less predictable motion relative to regions of varied sensitivity on the spacecraft detector, generating more noise in transiting exoplanet light curves. Using our simulations, we demonstrate that current de-trending techniques  effectively capture and remove systematics caused by sensitivity variation for spacecraft motion as high as about ten times that typically experienced by \textit{K2}. The \texttt{scope} package is open-source and has been generalized to allow custom detector and stellar target parameters. Future applications include simulating observations made by the Transiting Exoplanet Survey Satellite (TESS) and ground based observations with synthetic atmospheric interference as testbeds for noise-removal techniques.

\end{abstract}


\section{Introduction}

Despite the failure of two reaction wheels in 2012 and 2013, the \textit{Kepler} Space Telescope produced valuable data in its new configuration, \textit{K2}, with significantly higher precision than ground based telescopes \citep{2014PASP..126..398H}. However, due to the unstable pointing caused by the missing reaction wheels, targets in \textit{K2} observations have significant motion relative to the quantum sensitivity variation of the telescope detector, creating noise in \textit{K2} light curves. A number of attempts have been made to isolate and remove the instrumental noise from \textit{K2} data \citep{2014PASP..126..948V,2015A&A...579A..19A, 0004-637X-806-1-30, 2015MNRAS.454.4159H, 2015MNRAS.447.2880A, 2016MNRAS.459.2408A}. Through the application of data processing pipelines, namely the EPIC Variability Extraction and Removal for Exoplanet Science Targets (\texttt{EVEREST}) pipeline developed by \cite{2016AJ....152..100L}, the noise in \textit{K2} light curves can be reduced to the level of the original \textit{Kepler} mission for up to $15^{\text{th}}$ magnitude \citep{2018AJ....156...99L}.

However, as the \textit{Kepler} Space Telescope ran out of fuel, its motion due to thruster fires became less predictable and the magnitude of targets' motion relative to the detector increased. With higher motion, targets traversed more regions of varied pixel sensitivity, contributing more noise to the light curves of \textit{K2} targets. This increase in motion also increased the likelihood of flux pollution from neighbors, as stellar PSFs were more likely to overlap over the course of a campaign. Larger apertures were required for targets observed during motion because more pixels were sampled, further contributing to flux pollution from nearby stars.

In this paper, we present an approach to characterize increased noise due to high motion and assess noise-removal methods. \S 2 describes the mathematical methods of our simulation, \S 3 explores the results of our testing, \S 4 presents discussion of the context and future applications of this work, and \S 5 discusses our conclusions. Usage examples can be found in Appendix A.

\section{Methods}

Effective removal of instrumental noise requires a thorough understanding of its source. Stellar motion relative to the pixel sensitivity variation on the \textit{Kepler} CCD causes fluctuation in the amount of light recorded by the telescope detector over time. In order to accurately simulate the instrumental noise characteristic of \textit{K2} light curves, it is necessary to generate a forward model for the pixel sensitivity variation of the CCD. The properties of simulated targets are well understood, allowing the output light curves to be used for a variety of analyses. In particular, we use the simulated data to characterize noise levels resulting from conditions of high spacecraft motion and test the efficacy of de-trending methods. To accurately represent the \textit{Kepler} CCD, we generated a model for the detector that included both inter-pixel sensitivity variation between pixels and intra-pixel sensitivity variation within each pixel.

\subsection{PSF Model}

A stellar PSF was generated with a characteristic two-dimensional Gaussian shape, which includes covariance between $x$ and $y$ dimensions to capture PSF distortion due to incident light aberration on the \textit{Kepler} detector. We define our model for $F(t)$, the total flux absorbed by the telescope detector as a function of time, to be
\begin{equation}
F(t)=\sum_{aperture} \iint_{pixel} [s(x,y)P(x,y)\tau (t)]dxdy,\\
\end{equation}
where $s(x,y)$ is the sensitivity variation function, modeled by a low order polynomial, and $P(x,y)$ is the PSF of the star, centered at $(x_0,y_0)$ with amplitude $A$. $\tau (t)$ is a simulated transit model. The pixel sensitivity model in the $i^{\text{th}}$ pixel is given by the bivariate polynomial
\begin{equation}
s_i(x,y) = b_i\sum_n a_{x,n}x^n\sum_m a_{y,m}y^m, \\
\end{equation}
with coefficients $a_{x,n}, a_{y,m}$ defined to result in a sensitivity map that peaks in the center of each pixel and falls off toward the edge. The coefficient $b_i$ defines the inter-pixel sensitivity variation for the $i^{\text{th}}$ pixel. The PSF model is given by a sum over two-dimensional Gaussians
\begin{equation}
P(x,y) = A \sum_m \frac{1}{2\pi\sigma_{x,m}\sigma_{y,m}\sqrt{1-\rho_m^2}} \text{exp}\left[ -\frac{1}{2(1-\rho_m^2)} \left( \frac{(x-x_{0,m})^2}{2\sigma_{x,m}^2} + \frac{(y-y_{0,m})^2}{2\sigma_{y,m}^2} - \frac{2\rho_m  (x-x_{0,m})(y-y_{0,m})}{\sigma_{x,m}\sigma_{y,m}} \right) \right]
\end{equation}
where $\sigma_x$, $\sigma_y$ are the standard deviations of the PSF in $x$ and $y$, and $\rho$ is the correlation coefficient between $x$ and $y$. The mathematical PSF model (Equation 3) represents the detector's ideal response to the incident stellar flux before taking into account variation in pixel sensitivity. This ``ideally-measured" PSF is multiplied by the sensitivity function (Equation 2) to model the flux ultimately received by the CCD.

The coefficients $a_{x,n}, a_{y,n}$ were determined to emulate the magnitude of sensitivity variation on the \textit{Kepler} CCD based on its contribution to the noise in \textit{K2} light curves. Sensitivity coefficients are defined as independent in $x$ and $y$ for consistency with the Kepler Instrument Handbook \citep{kepler_intrument_handbook}. Our method for choosing values for sensitivity variation coefficients $a_n$ is described in the following section.

\subsection{Pixel Sensitivity Variation}

Our model for the sensitivity variation was chosen to emulate the same noise magnitude as real \textit{K2} observations. For our noise metric, we use the Combined Differential Photometric Precision (CDPP) \citep{10.1086/668847}. We used a two-step benchmarking process to estimate the magnitude of variation in quantum sensitivity and contribution by photon noise and background noise.

In Benchmark Test 1, we consider the no-motion case. We generate a set of high-precision simulated targets, which do not have increased systematic noise caused by motion. For this simulated data set, the noise versus magnitude trend should closely follow that of the original \textit{Kepler} mission. To most accurately capture the noise characteristics of the detector, our model included two sources of synthetic noise -- photon noise and background noise -- which contribute to the trend of increased noise for higher magnitude targets for the \textit{Kepler} mission. Benchmark Test 1 serves to constrain the levels of injected photon noise and background noise. The test consisted of plotting the CDPP for 1,000 targets from the original \textit{Kepler} mission and calculating the median value for each $Kp$ Mag at intervals of 0.5.

\begin{figure}[h]
	\centering
	\includegraphics[width=1.0\linewidth]{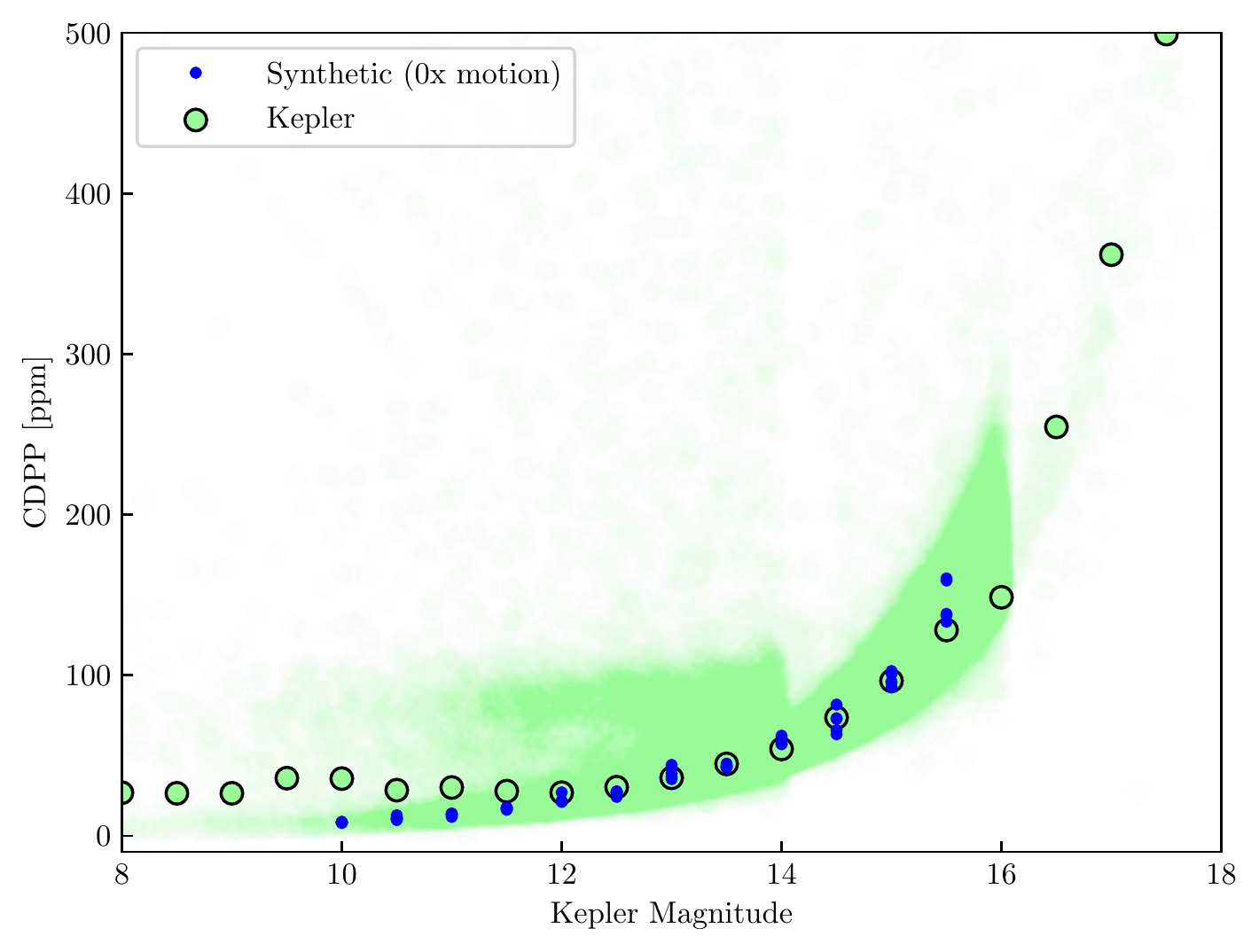}
	\caption{Benchmark Test 1: Noise level (CDPP) of our simulated light curves with no motion compared to original \textit{Kepler} targets as a function of $Kp$ Mag. The green trend demonstrates the relationship between $Kp$ Mag and noise for the \textit{Kepler} detector with high pointing precision, with the median represented by the larger green circles outlined in black. Each blue point is a simulated light curve with no motion vectors injected into the centroid position of the PSF. The primary contributors to this trend are photon noise and background noise, which were calibrated in our simulation by matching the slope of the \textit{Kepler} trend.}
	\label{fig:nomotion}
\end{figure}

We generated five synthetic stellar targets for each 0.5-Mag interval with the following random variation of PSF parameters for each individual simulation: centroid position varied by up to 0.2 pixels in both $x$ and $y$, $\sigma_x$ and $\sigma_y$ varied up to 10\% (${\sim}0.05$ pixels), and correlation coefficient $\rho$ varied up to 40\%. Each of these five simulated light curves is plotted against the \textit{Kepler} noise trend. This test was run iteratively on varied values of both background and photon noise until the slopes of the two trends approximately corresponded. To see the comparative noise of the resultant simulation, see Figure \ref{fig:nomotion}.

In Benchmark Test 2, we consider the case of typical \textit{K2} motion. To accomplish this, we injected motion from 1,000 cadences of the \textit{K2} target EPIC 205998445, creating a simulated example of a current observation to benchmark our model against \textit{K2} data. EPIC 205998445 is a \textit{K2} C03 star with $Kp = 12.029$ located ${\sim}60\%$ of the distance from the center of the detector plane to its edge. Statistics about the motion data used in our simulation can be found in Table \ref{table:motionstatistics}. This star was chosen because it is an isolated $12^{\text{th}}$ magnitude star, and the peak of the magnitude distribution for C03 was roughly 12. It is also located far enough from the center of the detector that it experiences significant motion during roll events, but not too near the edge that its motion is uncharacteristically high for \textit{K2} targets.

To calculate motion, the column pixel correction (header keyword \texttt{POS\_CORR1} in \textit{K2} fits files) and row pixel correction (header keyword \texttt{POS\_CORR2}), corresponding to motion in $x$ and $y$ respectively, were accessed from the target pixel file header. These offsets were added to the calculated centroid position of the simulated PSF for each cadence. Total magnitude of motion as reported in Table \ref{table:motionstatistics} represents the calculated maximum distance between centroid positions over the course of the simulated cadences.

\begin{figure}[h]
	\centering
	\includegraphics[width=1.0\linewidth]{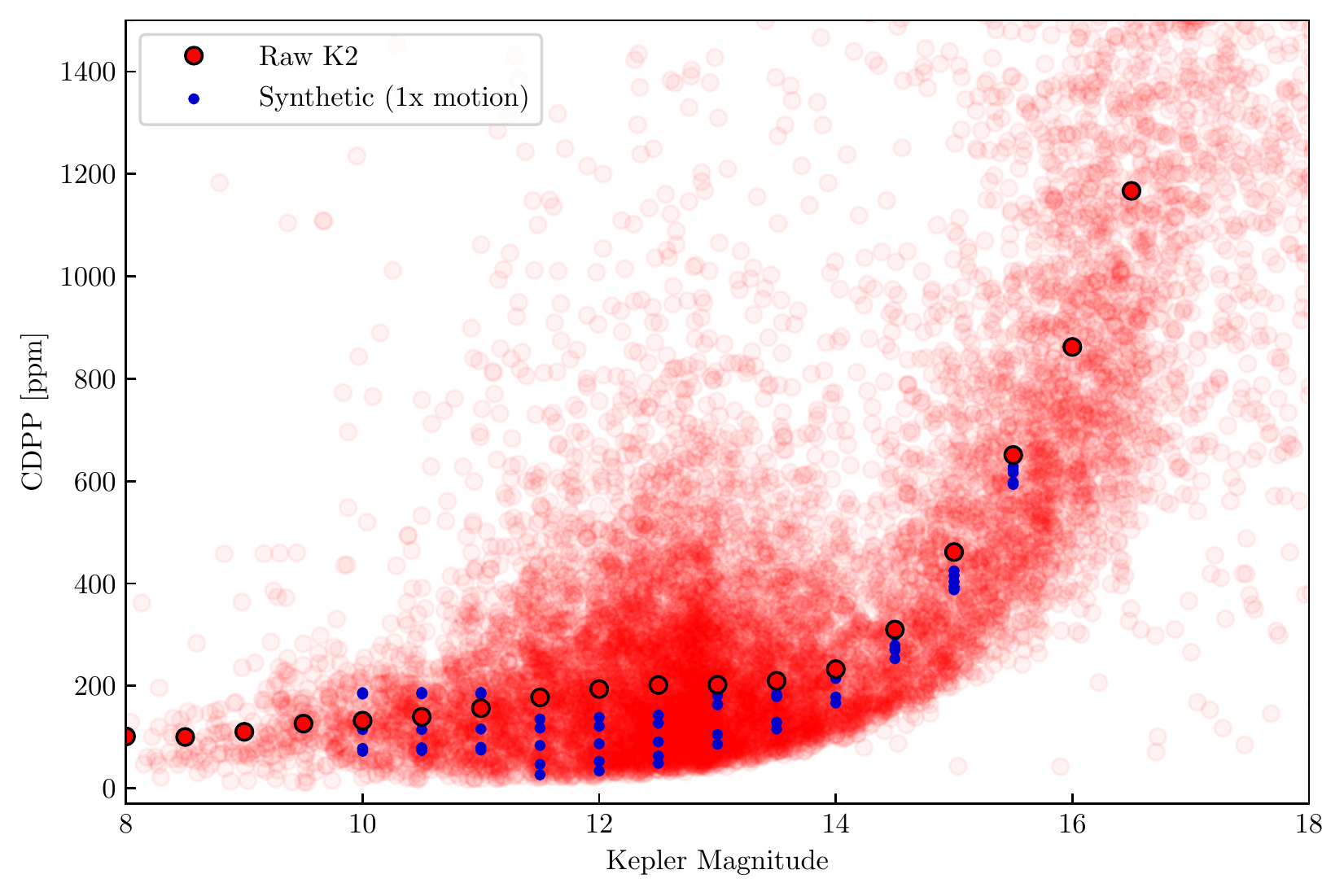}
	\caption{Benchmark Test 2: Noise level (CDPP) of our simulated light curves with \textit{K2} motion compared to real \textit{K2} targets as a function of $Kp$ Mag. The points displayed in red show the noise of 1,000 raw \textit{K2} observations before de-trending with the EVEREST pipeline, and the larger red points outlined in black follow the median of the trend. Each blue point is a simulated target with typical \textit{K2} motion and randomly varied initial centroid position and PSF shape. Our simulated data follow the primary trend of the CDPP vs. $Kp$ Mag relationship as real data. Where the simulated trend deviates roughly between $11^\text{th}$ and $13^\text{th}$ magnitude, the raw \textit{K2} CDPP is inflated for two reasons: first, the K2 observations contain a population of targets near the edge of the detector which have higher levels of systematic noise. A number of the observed targets also have short-period variability, which drives up CDPP levels. Our simulated targets are not placed at the edge of the detector and have no variability, and therefore follow the lower trend. This figure indicates that the sensitivity variation in our model appropriately captures the features of the detector.}
	\label{fig:1motion}
\end{figure}

With real \textit{K2} motion applied, sensitivity variation parameters were adjusted until CDPP versus $Kp$ Mag were  approximately  equal to the trend of real \textit{K2} observations. Sensitivity variation was altered by adjusting the $a_i$ coefficients in the polynomial model (Equation 2) to achieve different drops in sensitivity from pixels' centers to their edges, as well as adjusting the random variation in sensitivity between pixels with the $b_i$ sensitivity term. We found that a stochastic distribution of sensitivity with up to $1\%$ variation between pixels and $2.48\%$ variation within pixels, from center to edge, resulted in Benchmark Test 2 following the most similar trend to the expected result. Our sensitivity polynomial went up to third order with coefficients \{1, 0, -0.05\}. The results of Benchmark Test 2 can be seen in Figure \ref{fig:1motion}.

\begin{figure}[h]
	\centering
	\includegraphics[width=1.0\linewidth]{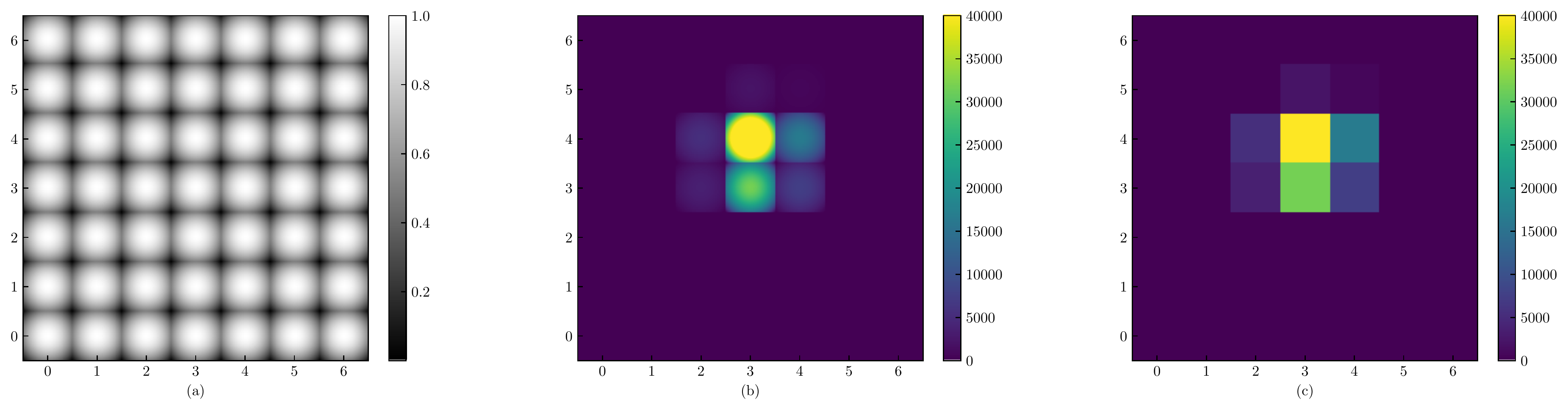}
	\caption{(a) A sample detector with sensitivity variation and no stellar targets, where white represents 100\% photon detection efficiency, and the shaded regions have lower values for quantum sensitivity. Note that our model peaks in sensitivity at the center of each pixel and falls off towards the edges. There is also slight random variation in sensitivity between pixels. (b) A Stellar PSF projected onto the detector at the sub-pixel level. (c) The final interpolated image. The value of each pixel is the integral over $x$ and $y$ of the sub-pixel flux. It should be noted that the sensitivity variation for this particular detector has been exaggerated to display the effects of sub-pixel response on the measurement of the PSF.}
	\label{fig:detector_sensitivity}
\end{figure}

A sample detector with included sensitivity variation can be seen in Figure \ref{fig:detector_sensitivity}.  For campaigns up to 17, stellar PSFs already traverse various regions of sensitivity variation on the CCD due to pointing instability. As spacecraft motion increased, PSFs moved more dramatically across many pixels, which contributed significant noise to sputtering-\textit{K2} light curves. With a working model for sensitivity variation, we examined the high-motion case.

\subsection{Increased Motion}

To test how our de-trending methods perform on targets with high motion relative to the detector, we injected motion vectors with various coefficients into our synthetic models. A coefficient of 1 corresponds to typical \textit{K2} motion (maximum of $\sim 0.59$ pixels), while a coefficient of 10 results in motion of up to $5.85$ pixels. More detail about the motion used in our simulations can be found in Table \ref{table:motionstatistics}.

\begin{table}[h!]
\begin{center}
    \begin{tabular}{c | c | c | c | c}
        Coefficient & Motion (pix) & Max Motion (pix) & Motion (arcsec) & Max Motion (arcsec) \\
        \hline \hline
        1 & $0.17\pm0.12$ & 0.59 & $0.67\pm0.47$ & 2.33 \\
        2 & $0.34\pm0.24$ & 1.17 & $1.34\pm0.95$ & 4.66 \\
				5 & $0.84\pm0.59$ & 2.93 & $3.34\pm2.37$ & 11.65 \\
				10 & $1.68\pm1.19$ & 5.85 & $6.68\pm4.74$ & 23.30 \\
   \end{tabular}
	 \caption{Coefficients applied to the motion vectors of \textit{K2} target EPIC 205998445, an isolated C03 star. A coefficient of 1 corresponds to typical motion, with each subsequent coefficient simply increasing the magnitude of motion by that factor. Motion is given in both pixels and arcseconds.}
	 \label{table:motionstatistics}
\end{center}
\end{table}

Our method to increase motion involved simply multiplying the pixel offset due to typical \textit{K2} motion by constant coefficients. This ensured the target traversed more pixels over a wider range of the detector while keeping the position calculation simple enough to avoid introducing new opportunity for error. It should be noted that this may not be the most appropriate treatment for all de-trending methods, particularly those that rely on fitting the centroid position of the target. However, the method used in the \texttt{EVEREST} pipeline (pixel level de-correlation) is agnostic to centroid position and relies simply on the measured flux in each pixel, so our treatment of motion is thorough enough for the goal of testing noise-removal methods.

With a forward model of \textit{K2} established, including both motion and sensitivity variation, we were able to apply our de-trending methods to stars with various tiers of increased motion to predict and understand the results of thruster failure. To elaborate on the results of our test, we first introduce our de-trending methods (\S 2.4), and then examine the application of these methods to our simulated data (\S 3.1).

\subsection{PLD}

After generating a set of light curves representative of extreme motion \textit{K2} observations, we tested systematics removal methods to assess the potential value of sputtering-\textit{K2} data. To de-trend our synthetic light curves, we used a variant of the method applied in the \texttt{EVEREST} pipeline. \texttt{EVEREST} utilizes a method called pixel level de-correlation (PLD), developed by \cite{0004-637X-805-2-132} for the \textit{Spitzer} Space Telescope.

We applied second-order PLD to our simulated light curves. Second-order PLD performed significantly better than the first-order model, but produced diminishing returns for higher orders when tested on our simulated observations with no variability. The second-order linear model for noise is defined by

\begin{equation}
m_i = \sum_l a_l \frac{f_{il}}{\sum_k f_{ik}} + \sum_l \sum_m b_{lm} \frac{f_{il}f_{im}}{\left( \sum_k f_{ik} \right)^2} + \alpha + \beta t_i + \gamma t_i^2
\end{equation}

where $m_i$ is the noise model at time $t_i$, $f_{il}$ is the flux in the $l^\text{th}$ pixel at time $t_i$, $a_l$ is the first-order PLD coefficient on the linear term, and $b_{lm}$ is the second-order PLD coefficient on the $l^\text{th}$, $m^\text{th}$ pixel pair. $\alpha$, $\beta$, and $\gamma$ are the Gaussian Process terms applied to capture long-period variability. Using Equation 4 to construct a model for the systematic noise, the remaining de-trending process is consistent with \cite{2016AJ....152..100L}.

PLD seeks to remove noise generated by intra-pixel sensitivity variation without solving for the centroid position of the target star. This method is particularly effective for \textit{K2} light curves despite the magnitude of apparent motion being high, and can recover \textit{Kepler}-like precision in exoplanet light curves for targets up to $Kp = 15$. For a more detailed treatment of PLD, see the paper by \cite{0004-637X-805-2-132} and the first two \texttt{EVEREST} papers \citep{2016AJ....152..100L,2018AJ....156...99L}.

\section{Results}

\subsection{Motion Tests}

\begin{figure}[h]
	\centering
	\includegraphics[width=1.0\linewidth]{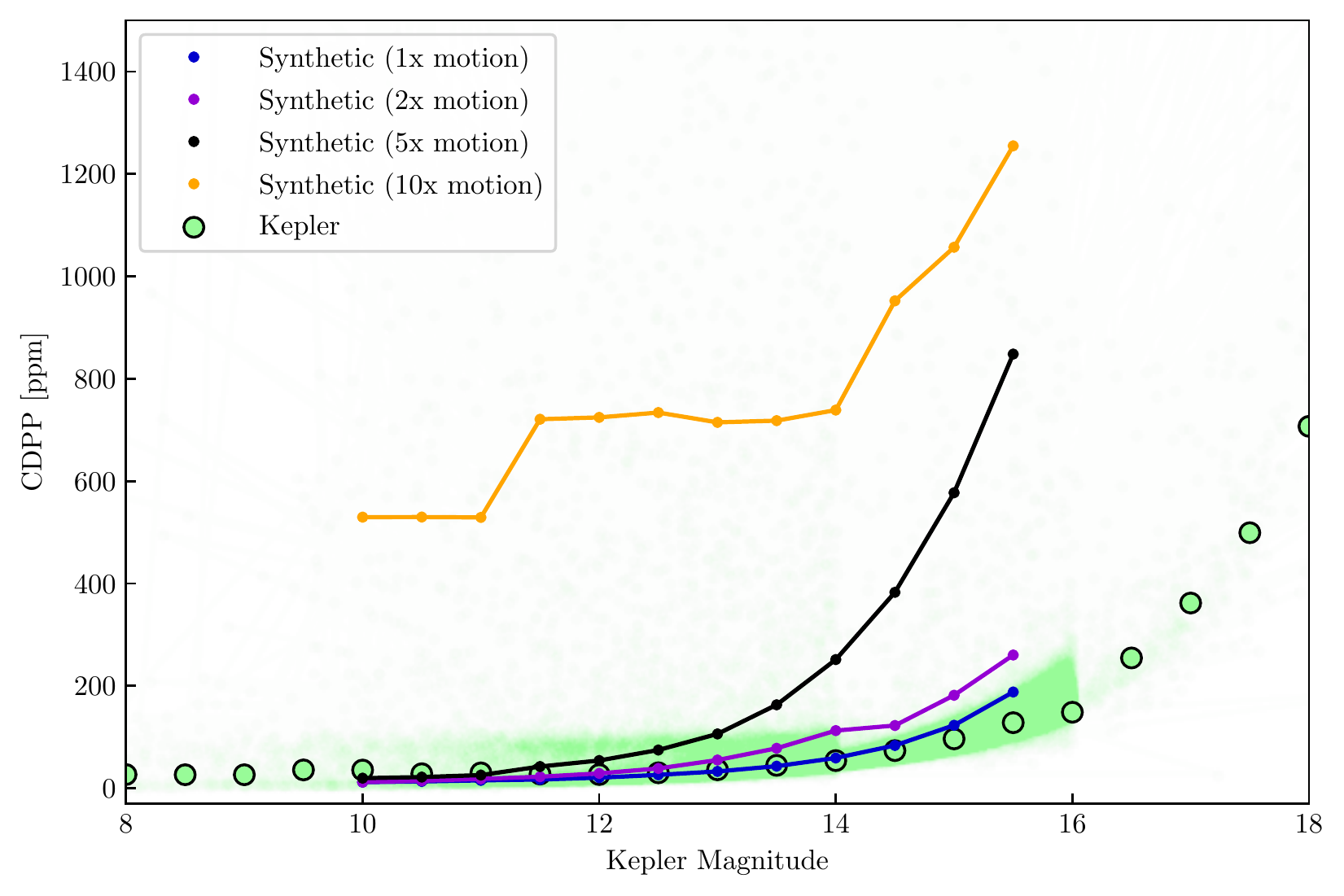}
	\caption{Noise level (CDPP) of simulated light curves after de-trending with $2^{\text{nd}}$ order Pixel Level De-correlation (PLD) as a function of $Kp$ Mag. The green points show noise levels of 1,000 targets from the original \textit{Kepler} mission with the median represented by the larger green circles outlined in black. De-trended simulated light curves with 1x, 2x, 4x, and 10x typical \textit{K2} motion are represented by the solid trends, and each point is the mean of 5 simulated light curves of a given $Kp$ Mag with slightly varied PSF shape and initial centroid positions. We achieve sub-400 ppm CDPP up to 5x motion for targets up to $Kp \text{ Mag}=14.5$, and de-trended light curves up to 10x motion perform better that ground-based observation (for which $<1,000$ ppm is difficult to achieve).}
	\label{fig:detmotion}
\end{figure}

For each motion coefficient, 5 simulated light curves were generated for magnitudes from $Kp$ Mag $=10$ to $Kp$ Mag $=15.5$ at 0.5-magnitude intervals. Each light curve was de-trended with $2^{\text{nd}}$ order PLD, and the CDPP was calculated. The average CDPP of the 5 simulations was calculated and plotted as a function of $Kp$ Mag. A detailed look at the results of our motion tests can be found in Figure \ref{fig:detmotion}.

Synthetic targets with 2x typical \textit{K2} motion achieved \textit{Kepler}-like photometry up to $Kp = 13$ when de-trended with $2^{\text{nd}}$ order PLD, and noise levels remained below 200 ppm up to $Kp = 15$ (compared to $<100$ ppm for \textit{Kepler}). Noise in light curves of targets with 5x motion remained below 400 ppm up to $Kp = 15.5$. After de-trending with PLD, we were able to retrieve transit parameters from simulated light curves with this magnitude of noise. This indicates that sustained observation during periods of increased spacecraft motion may result in the continued collection of valuable data.

These results were achieved by applying PLD to pixels within an aperture around our simulated stars. The aperture is defined to exclude extraneous background and photon noise in order to maximize light from our target. In applications for real targets, it is often necessary to also consider nearby bright stars when defining an aperture. The \texttt{scope} software package includes the ability to add neighboring stars, however they were excluded from these tests for simplicity. Future consideration will be given to automating aperture definitions, particularly because high motion PSFs will traverse more pixels and will therefore have a higher likelihood of crowding.

\section{Discussion}

Our simulations reproduce the noise properties of \textit{K2} observations and allow us to model potential sources of increased systematic noise. We have shown that modern de-trending methods, specifically PLD, remain effective at removing instrumental noise for observations taken during spacecraft thruster failure. In the following section, we demonstrate that our de-trended light curves with up to 10x typical motion retain higher precision than typical ground-based observations.

\subsection{Comparison to Ground-Based Photometry}

Ground-based photometric time series observations face challenges in achieving high precision due to atmospheric interference. \cite{Stefansson_2017} presents a comparison of modern high-precision observations from the ground. Their comparison reports that telescopes with  diameters similar to that of \textit{Kepler} fail to achieve sub parts-per-thousand precision from the ground. Observations with modern methods such as adaptive optics, defocusing, and defusers can achieve sub-300 ppm precision with telescopes of diameter greater than 3.5 m.

For simulated observations with with up to 2 times typical motion, \textit{Kepler}-like photometry is still possible up to $Kp=13$. For up to 10 times typical motion, sub-1,000 ppm is possible to achieve, performing better than ground-based observations for similar telescope diameters.

\subsection{Direct Measurement of Sensitivity}

A recent publication by \cite{2018arXiv180607430V} measured a dramatic drop in the sensitivity of \textit{Kepler} CCD pixels from center to edge. We acknowledge the inconsistency in resultant intra-pixel sensitivity variation between our calibration tests and the direct measurement of the sub-pixel response. Future work is necessary to identify the source of the meaningfully different conclusions, and we would like to identify a number of factors that likely contribute. The primary suspect is the degeneracy between inter- and intra-pixel sensitivity variation in our motion benchmark, both of which contribute to the systematic noise of synthetic light curves. Also included in this degeneracy are the magnitudes of background noise and photon noise. With some variation in the distribution of sensitivity, it is possible to achieve similar noise values as those estimated for \textit{K2} observations. Our values were determined by calculating the best fit of our synthetic light curves to observed noise trends.

Additionally, our calibration was based on \textit{K2} observations, and there are a number of physical differences between the state of the space telescope detectors and the laboratory environment. \cite{2018arXiv180607430V} acknowledge the temperature difference, and when considered in combination with the potential degradation of the telescope detector after nearly a decade of use, this could result in a significant difference in measured sensitivity.

It is also possible that the size and brightness of a stellar PSF would elicit a different response from the CCD's pixels compared to the much smaller point-scan PSF used to measure sub-pixel response. \cite{1748-0221-9-03-C03048} demonstrate that with increased brightness, the spread of a PSF also increases more than would be expected by a linearly increasing PSF, possibly due to the repulsive forces between electrons captured in a pixel. Because of this effect, the light curve generated by performing aperture photometry on larger, brighter PSFs may produce a level of noise consistent with a less significant drop intra-pixel sensitivity variation.

\subsection{Future Applications}

The Transiting Exoplanet Survey Satellite (TESS) recently started delivering light curves for nearby stars. TESS pixels are larger than those of \textit{Kepler} (21 arcseconds per pixel compared to \textit{Kepler}'s 3.98), and contamination by nearby targets is common. In the instance that TESS also faces motion issues, there will be a significant need to test de-trending methods with an emphasis on defining ideal apertures. With a combination of crowded PSFs and increased noise, apertures and associated aperture-based PLD methods will be an essential piece of our noise-removal strategy for TESS.

Further, the space telescope's trajectory could cause a periodic variation in the incident sunlight on the telescope, which will affect the quantum sensitivity of pixels. Periodically changing sensitivity could cause TESS PSFs to ``breathe" as the temperature rises and falls, and apertures around TESS targets will need to expand and contract correspondingly to effectively capture the desired stellar signal. Our forward model serves as an established simulation for cases of time-variable PSFs due to both motion and periodic sensitivity changes. The stellar PSF parameters can be easily adjusted in the forward model produced by \texttt{scope} to cover more pixels and more accurately capture PSFs characteristic of TESS observations.

Seeing aberration and atmospheric interference are additional sources of noise that could be included in the \texttt{scope} model. PLD was developed for \textit{Spitzer} and modified for application to \textit{K2}, but has seen little application elsewhere. For ground-based observations with significant noise due to pointing issues or atmospheric interference, PLD may be a valuable tool to make the most of data. \texttt{scope} is an ideal testbed for this type of noise-removal because it allows users to perform tests on noisy observations of synthetic sources that are well understood with input parameters that can be compared to post-processing results. Though it was initially imagined for space-based telescopes, this software has potentially useful applications for ground-based telescopes.

\section{Conclusions}

We have created a forward model of the \textit{Kepler} space telescope detector, \texttt{scope}, which includes inter- and intra-pixel sensitivity variation and mathematical simulations of stellar PSFs to test de-trending methods for exoplanet targets. Using these simulations, we have demonstrated that PLD is capable of reducing systematic noise to higher precision than ground-based observations in \textit{K2} light curves with up to 10x typical spacecraft motion. We have shown that in the event of spacecraft thruster sputtering due to diminishing fuel reserves, valuable data can likely still be collected by space telescopes. This provides valuable insight for the future of space-based observatories such as TESS, CHEOPS, and JWST in the event that they experience similar systematics to \textit{K2}. The question surrounding the usefulness of high-motion data is increasingly relevant to the study of \textit{K2}'s observations, and our hope is to emphasize the potential value still untapped in data collected during the final campaigns of observation.

Though TESS, a successor to \textit{Kepler}, has launched, there remains a wealth of existing archival \textit{K2} data that offers valuable contribution to our understanding of exoplanetary systems. Earth-like exoplanets can be difficult to identify in the high-magnitude noise of \textit{K2} data, which continued to get noisier as the mission reached its conclusion. TESS and JWST will likely face their own challenges with data analysis and noise-removal, and the tools developed for \textit{K2} will be foundations and references for new solutions. The methods we have presented aim to produce the most precise exoplanet identification, characterization, and analysis from the data we have available, and prepare for the next generation of space telescopes.

A fixed version of the open source \texttt{Python} package \texttt{scope} \citep{saunders_scope} can be found at

\texttt{doi:10.5281/zenodo.2542227}.

Follow the development or make contributions at \href{https://github.com/nksaunders/scope}{github.com/nksaunders/scope}.

This work was supported by NASA grant NNX14AK26G and by the NASA Astrobiology Institute's Virtual Planetary Laboratory. We would like to thank Geert Barentsen, Christina Hedges, and Michael Gully-Santiago for valuable discussion. Computing for this research was performed on the Hyak supercomputer system at the University of Washington.

\software{scope \citep{saunders_scope}, AstroPy \citep{astropy:2013, astropy:2018}, celerite \citep{celerite}, Everest \citep{2016AJ....152..100L}, Matplotlib \citep{Hunter:2007}, NumPy \citep{numpy}, SciPy \citep{scipy}, starry \citep{luger_starry}.}

\clearpage
\bibliographystyle{aasjournal}
\bibliography{references.bib}

\clearpage
\appendix
\section{Using \texttt{scope}}

All of our code is open-source and publicly available online. Our simulation package can be installed by running
\begin{lstlisting}[language=bash]
pip install tele-scope
\end{lstlisting}
To create a light curve with \texttt{scope}, first instantiate a \texttt{Target} object by calling:
\begin{lstlisting}[language=Python]
import scope
star = scope.generate_target()
\end{lstlisting}
An array of images (\texttt{targetpixelfile}), a one-dimensional flux light curve (\texttt{lightcurve}), and an array of matrices containing the error in each pixel (\texttt{error}) are all accessible as properties of the \texttt{Target} object:
\begin{lstlisting}[language=Python]
star.targetpixelfile, star.lightcurve, star.error
\end{lstlisting}
After a target is created, transits, variability, or neighbors can be added with the following functions (with example parameters provided):
\begin{lstlisting}[language=Python]
star.add_transit(depth=.01, per=15, dur=.5, t0=5.)
star.add_variability(var_amp=0.005, freq=0.25)
star.add_neighbor(magdiff=1., dist=1.7)
\end{lstlisting}
Alternatively, a \texttt{Target} can be instantiated containing added features with default parameters by changing the following boolean parameters:
\begin{lstlisting}[language=Python]
star = scope.generate_target(transit=True, variable=True, neighbor=True)
\end{lstlisting}
The detector with sensitivity variation can be displayed by calling
\begin{lstlisting}[language=Python]
star.display_detector()
\end{lstlisting}
To de-trend with second order PLD, run
\begin{lstlisting}[language=Python]
star.detrend()
\end{lstlisting}

Documentation and examples can be found online at \href{https://nksaunders.github.io/scope}{nksaunders.github.io/scope}. The code is publicly available at
\href{https://github.com/nksaunders/scope}{github.com/nksaunders/scope}.

\end{document}